\documentclass[journal=inoraj,manuscript=article]{achemso}

\usepackage{graphicx}
\usepackage{dcolumn}
\usepackage{bm}
\usepackage{xr-hyper}
\usepackage{hyperref}
\usepackage{booktabs}

\usepackage{xcolor} 

\SectionNumbersOn

	\title{Displacive Jahn--Teller transition in NaNiO$_2$}

	\author{
		Liam A. V. Nagle-Cocco*
		\footnote{Email: lavn2@cam.ac.uk.}~
	}
	\affiliation{Cavendish Laboratory, University of Cambridge, JJ Thomson Avenue, Cambridge, CB3 0HE, United Kingdom.}
	\email{lavn2@cam.ac.uk}
	
	\author{Annalena R. Genreith-Schriever}
	\affiliation{Yusuf Hamied Department of Chemistry, University of Cambridge, Cambridge, CB2 1EW, United Kingdom.}
	
	\author{James M. A. Steele}
	\affiliation{Cavendish Laboratory, University of Cambridge, JJ Thomson Avenue, Cambridge, CB3 0HE, United Kingdom.}
	\alsoaffiliation{Yusuf Hamied Department of Chemistry, University of Cambridge, Cambridge, CB2 1EW, United Kingdom.}
	
	\author{Camilla Tacconis}
	\affiliation{Cavendish Laboratory, University of Cambridge, JJ Thomson Avenue, Cambridge, CB3 0HE, United Kingdom.}
	
	\author{
		Joshua D. Bocarsly
		\footnote{Present address: Department of Chemistry, University of Houston, Houston, Texas, 77204-5003, United States of America.}~
	}
	\affiliation{Cavendish Laboratory, University of Cambridge, JJ Thomson Avenue, Cambridge, CB3 0HE, United Kingdom.}
	\alsoaffiliation{Yusuf Hamied Department of Chemistry, University of Cambridge, Cambridge, CB2 1EW, United Kingdom.}
	
	\author{Olivier Mathon}
	\affiliation{European Synchrotron Radiation Facility, 38043 Grenoble Cedex 9, France.}
	
	\author{Joerg C. Neuefeind}
	\affiliation{Spallation Neutron Source, Oak Ridge National Laboratory, Oak Ridge, TN 37831, United States of America.}
	
	\author{Jue Liu}
	\affiliation{Spallation Neutron Source, Oak Ridge National Laboratory, Oak Ridge, TN 37831, United States of America.}
	
	\author{Christopher A. O'Keefe}
	\affiliation{Yusuf Hamied Department of Chemistry, University of Cambridge, Cambridge, CB2 1EW, United Kingdom.} 
	
	\author{Andrew L. Goodwin}
	\affiliation{Inorganic Chemistry Laboratory, Department of Chemistry, University of Oxford, Oxford, OX1 3QR, United Kingdom.}	
	
	\author{Clare P. Grey}
	\affiliation{Yusuf Hamied Department of Chemistry, University of Cambridge, Cambridge, CB2 1EW, United Kingdom.}
	
	\author{John S. O. Evans}
	\affiliation{Department of Chemistry, Durham University, Durham, DH1 3LE, United Kingdom.}
	
	\author{
		Si\^an E. Dutton*
		\footnote{Email: sed33@cam.ac.uk.}~
	}
	\affiliation{Cavendish Laboratory, University of Cambridge, JJ Thomson Avenue, Cambridge, CB3 0HE, United Kingdom.}
	\email{sed33@cam.ac.uk}

\begin{document}
	
	
	\begin{abstract}
		Below its Jahn--Teller transition temperature, $T_\mathrm{JT}$, NaNiO$_2$ has a monoclinic layered structure consisting of alternating layers of edge-sharing NaO$_6$ and Jahn--Teller-distorted NiO$_6$ octahedra.
		Above $T_\mathrm{JT}$ where NaNiO$_2$ is rhombohedral, diffraction measurements show the absence of a cooperative Jahn--Teller distortion, accompanied by an increase in the unit cell volume. 
		Using neutron total scattering, solid-state Nuclear Magnetic Resonance (NMR), and extended X-ray absorption fine structure (EXAFS) experiments as local probes of the structure we find direct evidence for a displacive, as opposed to order-disorder Jahn--Teller transition at $T_\mathrm{JT}$. 
		This is supported by \textit{ab initio} molecular dynamics (AIMD) simulations. 
		To our knowledge this study is the first to show a displacive Jahn--Teller transition in any material using direct observations with local probe techniques.
	\end{abstract}

	\maketitle

	\section{\label{sec:level1}Introduction}
	
	In an undistorted octahedrally-coordinated environment, partially-occupied $e_g$ orbitals in a transition metal cation would be degenerate and therefore unstable to symmetry-reducing distortions by the Jahn--Teller (JT) effect~\cite{jahn1937stability}. 
	This lifts the degeneracy of both the $t_{2g}$ and partially-occupied $e_g$ orbitals, lowering the energy of the system~\cite{jahn1937stability,opik1957studies,longuet1958studies,goodenough1998jahn}. 
	The resulting octahedral distortion is often a linear combination of two possible van Vleck modes~\cite{van1939jahn,naglecocco2023van}: a planar rhombic $Q_2$ distortion and a tetragonal $Q_3$ elongation/compression [Figure~S1]. 
	Experimentally, the distortion is normally dominated by the $Q_3$ tetragonal elongation. 
	This JT distortion, and associated orbital ordering, is relevant to many phenomena including unconventional superconductivity in the cuprates~\cite{fil1992lattice,keller2008jahn}, magnetic structure through coupling of spin and orbital ordering~\cite{khomskii2020orbital}, and ionic mobility~\cite{kim2015anomalous,li2016jahn}, and can lead to structural transitions in a material when it is used in a battery electrode~\cite{kim2015anomalous,li2016jahn,choi2019k0}. 
	
	\begin{figure*}
		\includegraphics[scale=1]{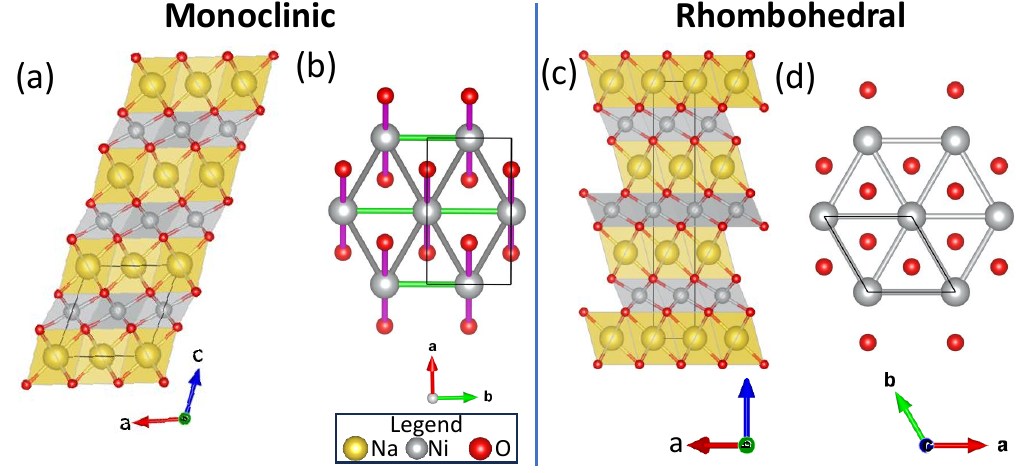}
		\caption{
			Crystal structures of (a) $C2/m$ monoclinic and (c) $R\bar{3}m$ rhombohedral NaNiO$_2$. 
			(b) and (d) show the Ni-Ni distances (grey for non-elongated; green for elongated) and JT-elongated Ni-O distances (pink) in these two structures. In the rhombohedral structure there are no elongated Ni-Ni or Ni-O bonds. 
		}
		\label{crystal-structures.pdf}
	\end{figure*}
	
	In a crystalline material, the energy of a JT-distorted system is often minimised when the axes of elongation of neighbouring octahedra are correlated; this is termed a cooperative JT distortion, in contrast to a non-cooperative system in which axes of elongation are randomly distributed~\cite{goodenough1998jahn}. 
	Transitions from low-temperature, cooperative JT distortions to a high-temperature state with an undistorted average structure can generally be classified as order-disorder or displacive~\cite{radin2020order}. In the former case, non-cooperative JT distortions persist locally but are averaged out in the bulk structure, whereas in the latter case the local structure is undistorted. 
	This paradigm has been applied to several systems~\cite{qiu2005orbital,souza2005local,thygesen2017local,radin2020order,genreith2023jahn}. 
	The most well-studied is in the JT-distorted $d^4$ Mn$^{3+}$ perovskite LaMnO$_3$ which shows evidence for an order-disorder JT transition~\cite{araya2001local,qiu2005orbital,garcia2005jahn,souza2005local}; theoretical works~\cite{ahmed2005phase,ahmed2006potts,ahmed2009volume} and total scattering experiments~\cite{thygesen2017local} indicate a transition to a Potts model~\cite{ahmed2005phase} type of orbital disorder. 
	
	The order-disorder/displacive paradigm has also been applied to the layered transition metal oxides with formula $AM$O$_2$ ($A$=alkali metal, $M$=transition metal).~\cite{radin2020order,genreith2023jahn} This family of materials includes several battery materials such as LiNiO$_2$, NaNiO$_2$, LiMnO$_2$, and NaMnO$_2$, all of which feature transition metal ions with degenerate $d^7$/$d^4$ $e_g$ orbitals which are liable to JT distortions. 
	The $AM$O$_2$ materials have layers of edge-sharing $M$O$_6$ octahedra forming a triangular network of $M$ cations, separated by a layer of octahedrally-coordinated alkali metal ions, in contrast to the perovskites such as LaMnO$_3$ which have corner-sharing octahedra and an approximately cubic cation network.
	The aristotype of the structure has rhombohedral $R\bar{3}m$ symmetry, but NaNiO$_2$ and each of $\alpha$- and $\beta$-$X$MnO$_2$ ($X$=Li, Na) show cooperative JT ordering~\cite{radin2020order} resulting in a macroscopic distortion. 
	NaNiO$_2$ exhibits a monoclinic $C2/m$ distortion due to collinear JT-elongated octahedra, which disappears on heating during a first-order transition which onsets at 460\,K~\cite{dyer1954alkali,dick1997structure,chappel2000study,sofin2005new,nagle2022pressure} (see Figure~\ref{crystal-structures.pdf}) to the aristotype $R\bar{3}m$ structure. 
	LiMnO$_2$ and NaMnO$_2$ exhibit polymorphism and more complex cooperative ordering~\cite{radin2020order}. 
	LiNiO$_2$ is a complicated case, with clear experimental evidence for multiple Ni-O bond lengths~\cite{pickering1993nickel,rougier1995non,chung2005local} but diffraction data are typically modelled with the aristotype structure which does not allow for cooperative JT ordering. 
	There have been a broad array of proposed forms of JT ordering~\cite{chung2005local,chen2011first} for LiNiO$_2$, with the energetically most-favourable being a zigzag ordering with monoclinic $P2_1/c$ symmetry~\cite{chen2011first,chen2011charge,radin2018simulating,foyevtsova2019linio,genreith2023jahn}. 
	Alternative phenomena for LiNiO$_2$ involving spin- or even charge-disproportionation~\cite{chen2011charge,foyevtsova2019linio,poletayev2022temperature} have also been proposed; these are likely a feature of other nickelates such as AgNiO$_2$~\cite{wawrzynska2007orbital,kang2007valence} or the nickelate perovskites~\cite{garcia1994neutron,mizokawa2000spin,garcia2009structure}. 
	In contrast to a recent theoretical study~\cite{radin2020order}, which concludes that the layered alkali metal transition metal oxides $AB$O$_2$ ($A$=Li, Na; $B$=Ni, Mn) should exhibit an order-disorder JT transition, we have recently found evidence for a displacive transition in LiNiO$_2$ using \textit{ab initio} Molecular Dynamics (AIMD) and variable-temperature X-ray diffraction (XRD)~\cite{genreith2023jahn}. 
	
	In this work, we have studied the variable-temperature properties of NaNiO$_2$. We use variable-temperature synchrotron X-ray diffraction, neutron total scattering, Ni K edge extended X-ray absorption fine structure (EXAFS), $^{23}$Na magic-angle spinning (MAS) solid state nuclear magnetic resonance (NMR) spectroscopy, and AIMD to study the changes in NaNiO$_2$ with a focus on the local Ni environment through the monoclinic$\rightarrow$rhombohedral transition at $\sim$460\,K where the cooperative JT distortion disappears. We find evidence of a displacive JT transition, in contrast to the majority of previous studies on Jahn--Teller transitions which report order-disorder transitions. 
	Taken together with previous results~\cite{nagle2022pressure,genreith2023jahn}, this suggests a broader conclusion that the JT transitions in layered triangular-lattice nickelates are different from those in the transition metal perovskites where local JT distortions are reported to persist in the high-temperature phase. 
	
	\section{\label{sec:level4}Results}

	\subsection{Variable-temperature synchrotron X-ray diffraction.}
	
	\begin{figure*}
		\includegraphics[scale=0.9]{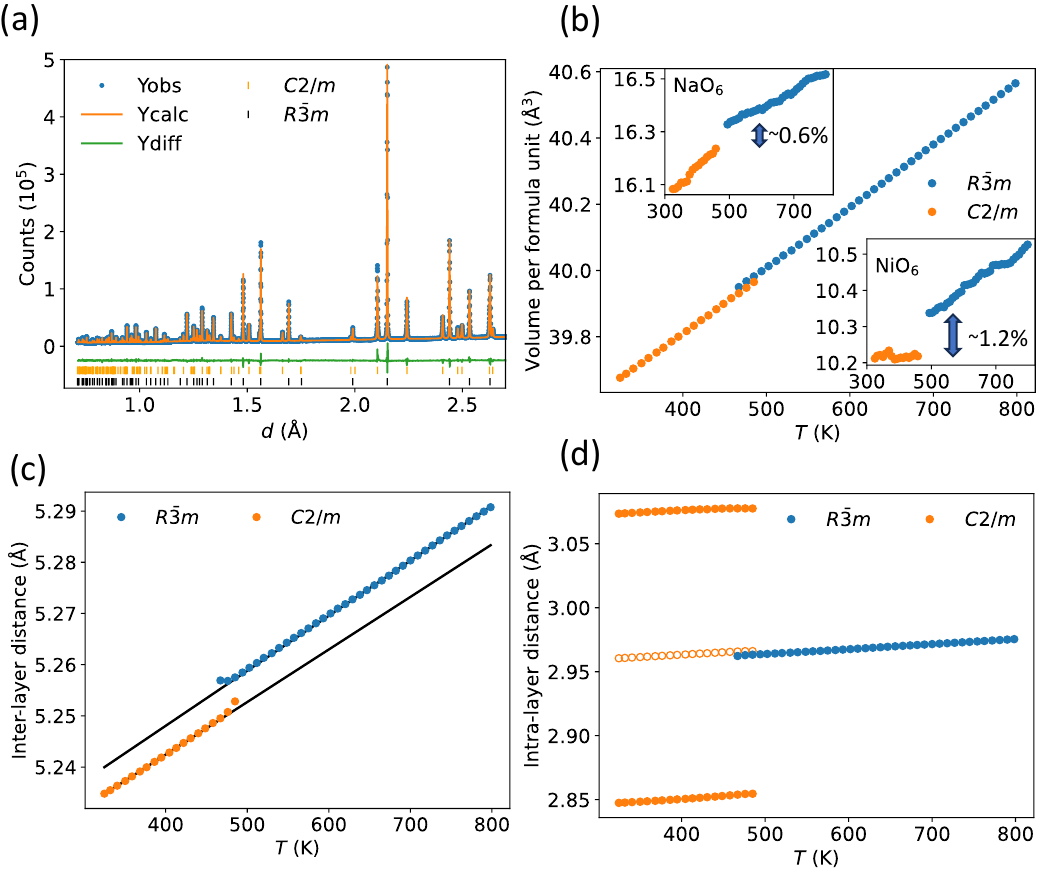}
		\caption{
			(a) Representative Rietveld refinement of synchrotron X-ray diffraction data for NaNiO$_2$ at 476.1\,K in the mixed-phase regime where the rhombohedral ($R\bar{3}m$) and monoclinic ($C2/m$) phases coexist. 
			(b) Volume per formula unit as a function of temperature, showing the slight increase in unit cell volume of the rhombohedral phase compared with the monoclinic phase. Insets are NaO$_6$ (top left) and NiO$_6$ (top right) octahedral volume with temperature, calculated using \textsc{VanVleckCalculator}; only volumes in the single-phase regions are plotted. 
			Temperature-dependence of the (c) inter-layer distances, $c/3$ for the rhombohedral phase and $c\sin{(\beta)}$ for the monoclinic phase, and (d) intra-layer distances, $a$ for the rhombohedral phase, and $a/\sqrt{3}$ and $b$ for the monoclinic phase. 
			The intra-layer distances correspond to the Ni-Ni and Na-Na distances within the plane. 
			For the monoclinic phase, closed circles are the individual distances, open circles are averaged. In (c), points are experimental values obtained by Rietveld refinement, and solid lines are a linear fit. In (b,c,d) error bars are smaller than data points.
		}
		\label{I11-synchrotron}
	\end{figure*}
	
	Variable-temperature synchrotron X-ray powder diffraction was performed to observe temperature-dependent changes in the average crystal structure and analysed by Rietveld refinement~\cite{rietveld1969profile}. 
	On heating, the crystal structure of NaNiO$_2$ is monoclinic ($C2/m$) until $\sim$460\,K where Bragg peaks associated with the high-temperature rhombohedral ($R\bar{3}m$) phase begin to emerge, indicating this is approximately the onset of the cooperative transition temperature $T_\mathrm{JT}$. 
	A mixed-phase regime exists, in which the phase fraction of the monoclinic phase decreases with heating, until the sample becomes entirely rhombohedral around 505\,K. 
	These findings are consistent with previous variable-temperature diffraction studies of NaNiO$_2$~\cite{dick1997structure,chappel2000study,sofin2005new}.

	Selected results of the Rietveld analysis are shown in Figure~\ref{I11-synchrotron}, with further information in SI Section~2. 
	The volume per formula unit is larger in the rhombohedral than the monoclinic structure [Figure~\ref{I11-synchrotron}(b)]. There is positive thermal expansion in both the monoclinic and rhombohedral phases. 
	In the monoclinic phase, NiO$_6$ octahedral volume is essentially temperature-independent [inset of Figure~\ref{I11-synchrotron}(b)] and the volume increase is entirely driven by NaO$_6$ octahedral expansion. 
	Through the phase transition, there is a discontinuous increase in unit cell volume per formula unit, primarily due to a $\sim$1.2\% jump in NiO$_6$ octahedral volume (compared with a $\sim$0.6\% increase for NaO$_6$ octahedra) from the fully monoclinic to fully rhombohedral phases. 
	In the rhombohedral phase both the NiO$_6$ and NaO$_6$ octahedra expand on heating at around the same rate. 
	
	Figure~\ref{I11-synchrotron}(c) shows that there is a significant increase in inter-layer distance on heating from the monoclinic to rhombohedral structures. 
	Figure~\ref{I11-synchrotron}(d) shows that the two different intra-layer distances (corresponding to Ni-Ni and Na-Na distances within the plane of the layer) of the monoclinic cell equalise in the rhombohedral cell, due to the increase in cell symmetry, and the distance in the rhombohedral cell is slightly decreased compared with the average distance in the monoclinic cell.
	
	\subsection{$^{23}$Na nuclear magnetic resonance.}

	A variable-temperature $^{23}$Na NMR (VT-NMR) experiment was carried out to investigate changes in the local structure with temperature using both magic angle spinning (MAS) and static measurements. 
	Changes in both of the spectra are observed at $T_\mathrm{JT}$ [Figure~S37]. 
	At low temperatures a single Na$^+$ environment is observed consistent with the average structure of monoclinic NaNiO$_2$. A second higher chemical shift environment is observed on heating which we ascribe to Na$^+$ in the high-temperature, rhombohedral phase of NaNiO$_2$. 
	There is a limited $T$-range where both peaks co-exist until at higher $T$ only a single peak persists. 
	Both environments have large hyperfine shifts due to the presence of paramagnetic Ni$^{3+}$ ions, the peak shifting to lower values as the Ni$^{3+}$ become less paramagnetic at higher temperatures. 
	These results are consistent with the diffraction data and indicate a change in the local Na$^+$ environment at $T_\mathrm{JT}$.

	\begin{figure*}
		\includegraphics[scale=0.8]{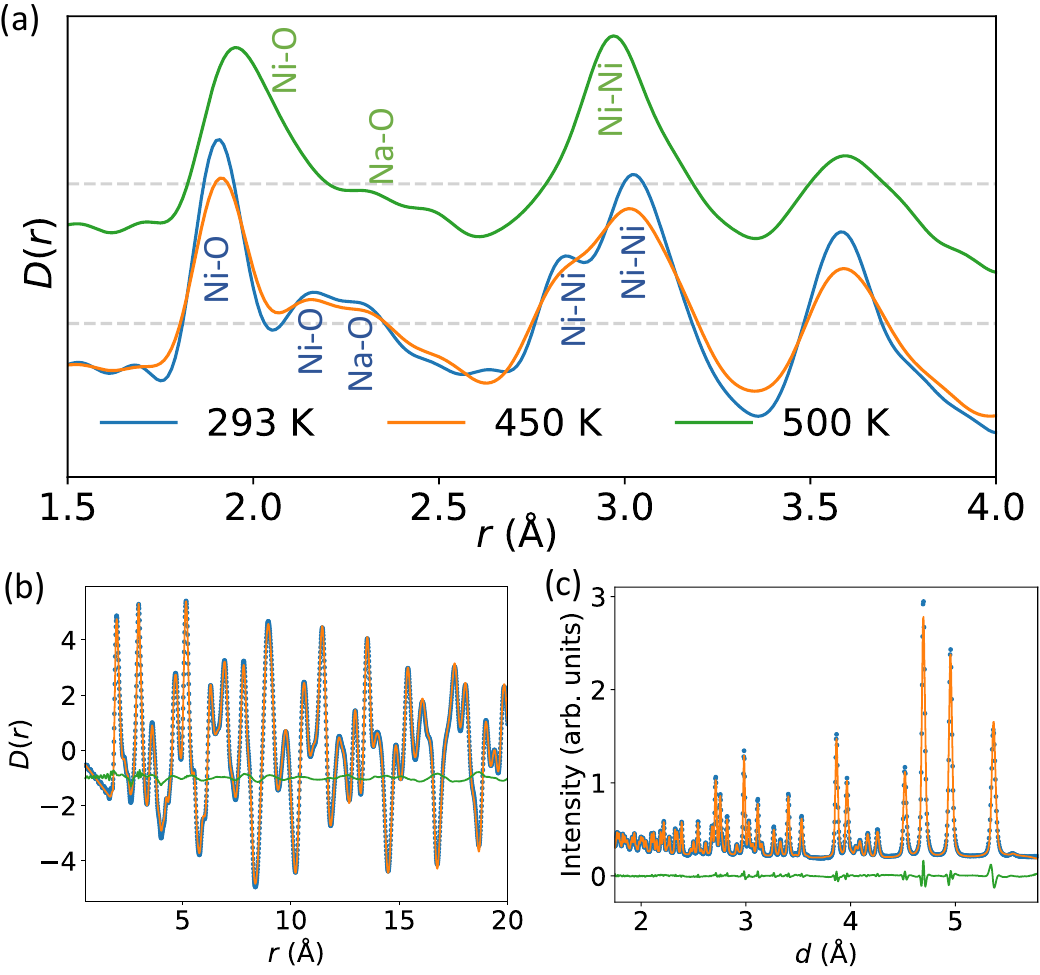}
		\caption{
			(a) Experimental neutron PDFs for NaNiO$_2$ at 293\,K, 450\,K, and 500\,K, in the $r$-range where the nearest-neighbour Ni-O peaks occur. 
			Peaks are labelled for the 293\,K dataset in blue text and 500\,K dataset in green text. 
			The 500\,K data is vertically offset from the low-temperature data for ease of distinguishing the changes compared with the lower-$T$ datasets. Grey dashed horizontal lines occur at $D(r)=0$ for the lower two temperatures (lower) and 500\,K (upper). 
			(b,c) Example fits to the PDF and Bragg diffraction data at 500\,K using the supercell fitting (experimental data: blue; fitted data: orange; difference pattern; green). 
		} 
		\label{PDF_fig1}
	\end{figure*}
	
	\subsection{Neutron pair distribution function.}
	
	Total scattering neutron experiments have been performed on NaNiO$_2$. The Bragg data was published previously~\cite{nagle2022pressure} and is consistent with the synchrotron XRD presented in the previous section. 
	
	Pair Distribution Functions (PDFs), presented as $D(r)$ in Figure~\ref{PDF_fig1}(a),~\cite{keen2001comparison} were obtained from the neutron total scattering data at 293\,K and 450\,K (both monoclinic average structure), and 500\,K (rhombohedral average structure). 
	There are two Ni-O peaks at 293\,K and 450\,K, consistent with the Ni-O bond length splitting due to a Jahn--Teller distortion, and a single Ni-O peak at 500\,K, implying that NiO$_6$ octahedra are undistorted by the Jahn--Teller effect at this temperature. 
	This is consistent with the picture from the average structure, as described from the variable-temperature synchrotron data. However, at 500\,K the Ni-O peak is highly asymmetric with a tail on the high-$r$ side. 
	
	Initial analysis of the PDF data was performed using small-box analysis, also known as real-space Rietveld refinement. 
	Results are presented in Section~S3.1 of the SI. 
	The room-temperature data can be fit well with the JT-distorted, monoclinic structure used for Rietveld analysis of the reciprocal space data. 
	The PDF data at 500\,K can be fit adequately with both the monoclinic and the rhombohedral (JT-undistorted) structures, with a slightly higher $R_\mathrm{wp}$ for the rhombohedral structure (although this is to be expected given the smaller number of free parameters). 
	In the monoclinic cell, the difference between the Ni-O bond lengths is finite but very small at 500\,K ($\sim$0.041\,\AA{}, compared with $\sim$0.243\,\AA{} at 293\,K), which is not consistent with the presence of a local Jahn--Teller distortion. 
	
	Ultimately the small-box analysis did not fit all features of the PDF data, and in any case small-box analysis with a single unit cell would be insensitive to some types of local Jahn--Teller ordering. 
	To test for the presence of local Jahn--Teller distortions obscured by the peak asymmetry, big-box analysis was performed on the PDF data, in conjunction with the Bragg scattering data. 
	
	To this end, a rhombohedral NaNiO$_2$ unit cell with centred orthorhombic setting was prepared, consisting of 6 formula units and obtained via transformation from the normal rhombohedral cell using Equation~S5. 
	A $16\times9\times3$ supercell of this unit cell was used as the starting point for big box refinements, with lattice dimensions $\sim$45\,\AA{} and containing 10368 atoms; see Methods for further details. 
	The atomic positions in this model were then refined against the experimental data. 
	Here, empirical bond valence sum~\cite{brown1985bond} restraints (discussed in Methods) were applied to the atomic positions to ensure physical behaviour, although qualitatively equivalent results are obtained without using these restraints [SI Section~S3.9]. The big box refinements were run several times to ensure repeatability [SI Section~S3.11]. 
		
\begin{figure}
		\includegraphics[scale=0.99]{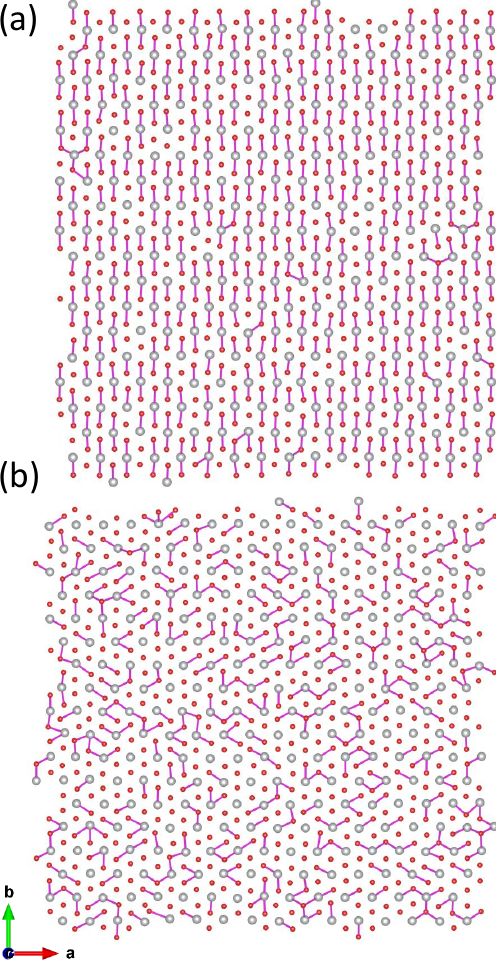}
		\caption{
			Cross sections of the $ab$ plane for a single O-Ni-O layer of the supercell obtained by big box PDF analysis, showing the distribution of elongated ($r>2.1$\,\AA{}) Ni-O bonds at (a) 293\,K and (b) 500\,K. Na ions are hidden for clarity. 
			Ni: grey; O: red.
		} 
		\label{PDF_Fig_CrossSections}
	\end{figure}

\begin{figure*}[t]
		\includegraphics[scale=0.8]{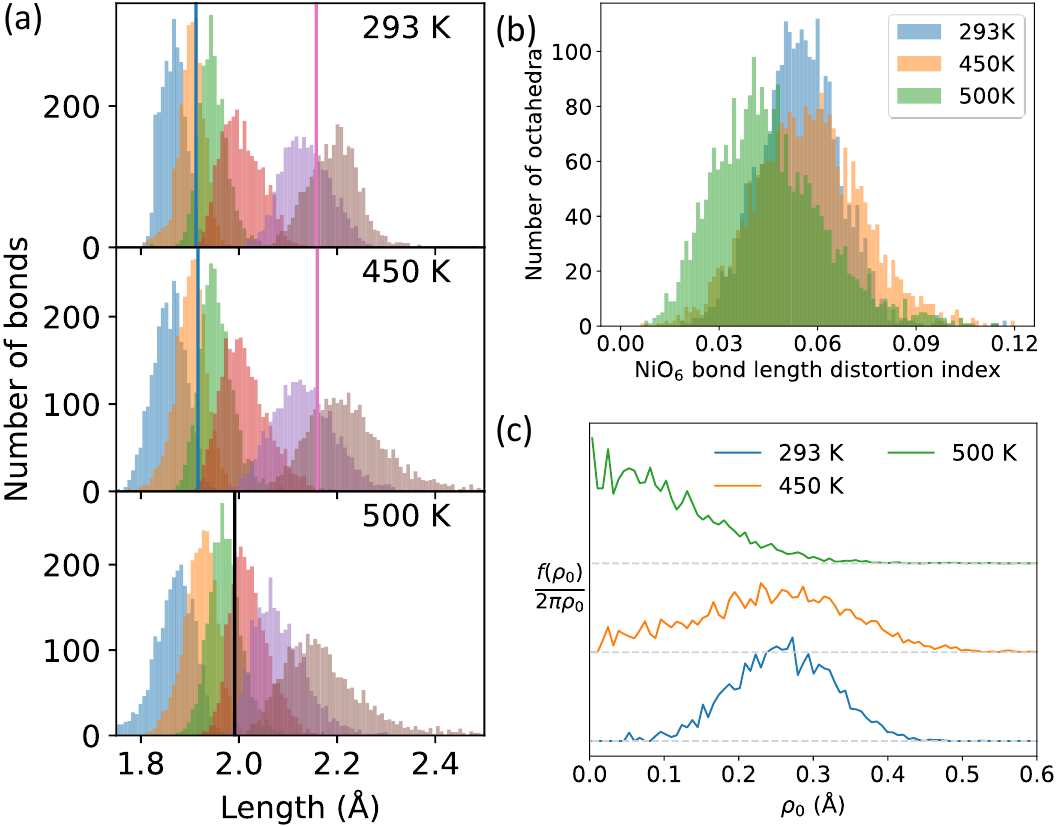}
		\caption{
			(a) Ni-O bond distributions at 293\,K (top), 450\,K (middle), and 500\,K (bottom), as obtained via big box analysis of neutron PDF data, for the smallest to largest Ni-O bond in each octahedron; vertical lines are the bond lengths from Rietveld refinement of the neutron data reported in Ref.~\cite{nagle2022pressure}. 
			(b) Histogram of bond length distortion index~\cite{baur1974geometry} for each NiO$_6$ octahedron at 293\,K, 450\,K, and 500\,K. 
			(c) Probability function of distortion [Equation~\ref{rho_0_prob}] at 293\,K, 450\,K, and 500\,K. 
			(b) and (c) are obtained using \textsc{VanVleckCalculator}~\cite{naglecocco2023van,VanVleckCalculator}.
		} 
		\label{PDF_fig2}
	\end{figure*}

	An example fit at 500\,K to both the reciprocal-space and real-space datasets can be seen in Figures~\ref{PDF_fig1}(b,c) and the resulting supercells are shown in Figure~\ref{PDF_Fig_CrossSections}. 
	Qualitatively, it can be seen by simple inspection of the O-Ni-O layers in the supercell that there is a cooperative $Q_3$ [Figure~S1] elongation of the NiO$_6$ octahedra at 293\,K (the Jahn--Teller distortion) which does not seem to be present at 500\,K. 
	These supercells are quantitatively analysed in terms of the Ni-O bond length distribution [Figure~\ref{PDF_fig2}(a)] and it can be seen that below the monoclinic$\rightarrow$rhombohedral transition there are two clusters of Ni-O bond lengths; a shorter cluster representing four bonds and a longer cluster with two bonds. 
	However, at 500\,K there is just a single cluster, which is not consistent with Jahn--Teller-distorted octahedra. 
	The model is consistent with the significant peak asymmetry in the Ni-O bond length observed in the PDF data, with the peak asymmetry occurring as a consequence of the broader bond length distribution for longer bonds. 
	
	Other evidence for the absence of Jahn--Teller distortion in the local structure of rhombohedral NaNiO$_2$ can be obtained by considering two metrics for the octahedral distortion. 
	The bond length distortion index (BLDI)~\cite{baur1974geometry}, Figure~\ref{PDF_fig2}(b), quantifies the deviation of bond lengths from the average for a given polyhedron and is often used to quantify Jahn--Teller distortions~\cite{kimber2012charge,lawler2021decoupling,nagle2022pressure}. 
	For a big box model where thermal effects are modelled by deviation of atoms from their average positions (rather than using atomic displacement parameters, as is conventional for Rietveld refinement), we would expect the BLDI to increase with heating in the absence of any electronic or magnetostructural transitions. 
	The average BLDI for all NiO$_6$ octahedra is 0.0547 and 0.0575 (to 3 significant figures) at 293\,K and 450\,K, respectively, in the monoclinic phase, before decreasing to 0.0458 at 500\,K. This is clearly shown by the decreasing distribution of bond lengths in Figure~\ref{PDF_fig2}(a).
	In contrast, the average BLDI continuously increases with heating for the NaO$_6$ octahedra, increasing from 0.0260 at 293\,K, to 0.0352 at 450\,K, and finally 0.0450 at 500\,K (see Figure~S10). 
	This trend in the BLDI for NiO$_6$ octahedra can be explained by the disappearance of a local Jahn--Teller distortion at $T_\mathrm{JT}$. 
	We can also quantify the distortion using the parameter $\rho_0$, where $\rho_0^2=Q_2^2+Q_3^2$. Figure~\ref{PDF_fig2}(c) shows $P(\rho_0)$, the probability of octahedra having a given value of $\rho_0$, defined as:
	
	\begin{equation} \label{rho_0_prob}
		P(\rho_0) = \frac{f(\rho_0)}{2\pi \rho_0}
	\end{equation}
	where $f(\rho_0)$ is a histogram of $\rho_0$ for all NiO$_6$ octahedra. 
	This parameterisation shows maximum probability at $\rho_0\approx0$ at 500\,K, compared with large non-zero magnitudes at 293\,K and 450\,K, which also indicates the absence of a local Jahn--Teller distortion in the rhombohedral phase. 
	This result that the most probable $\rho_0$ substantially decreases at 500\,K, compared with lower temperatures, is resilient to other ways of running the big box refinement, for instance with different starting configurations [SI Section~3.7], changes in the BVS restraints [SI Section~S3.9], and multiple runs with the same starting parameters [SI Section~3.11]. 
		
	\subsection{X-ray absorption spectroscopy.}
	
	\begin{figure*}
		\includegraphics[scale=0.8]{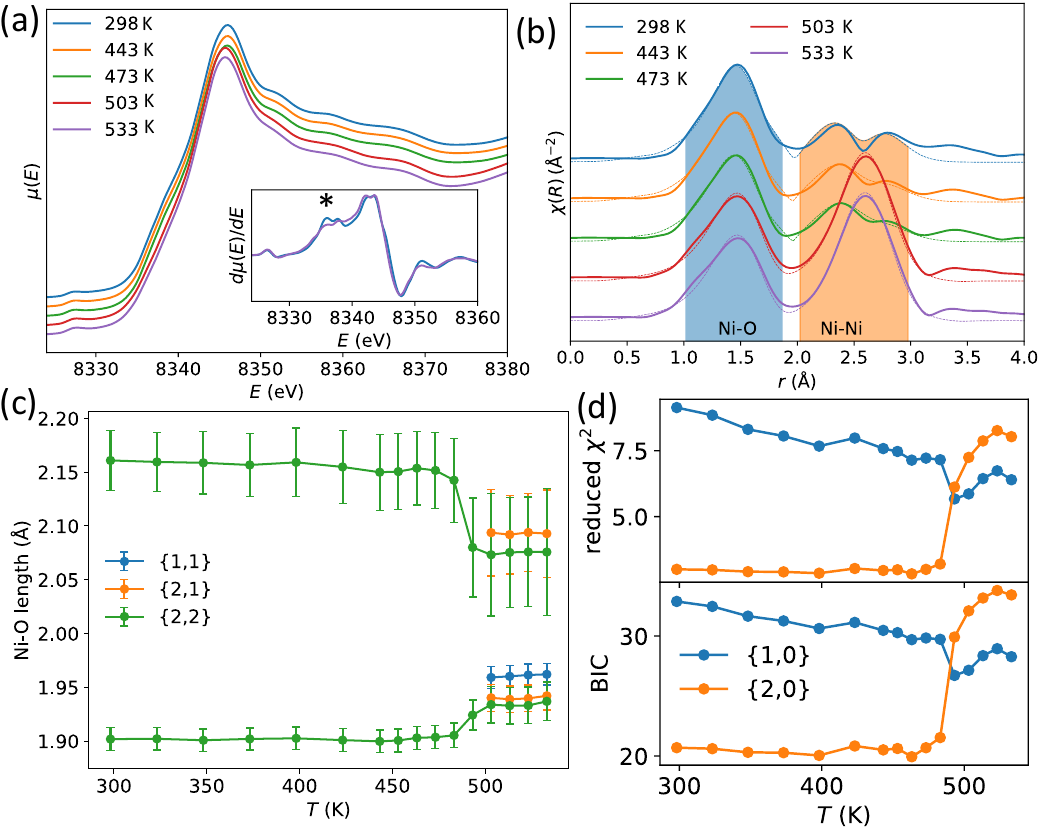}
		\caption{
			(a) XANES at the Ni K edge at various temperatures, where each spectrum is offset by an arbitrary amount for clarity; inset shows $d\mu/dE$ for the 298.15\,K and 533.15\,K, with an asterisk marking the feature on the rising edge. 
			(b) EXAFS data at the Ni K edge, with heating, for NaNiO$_2$ at various temperatures, with the solid line representing the fit with a \{2,2\} model consisting of two Ni-O bond lengths and two Ni-Ni interatomic distances. The feature around 3.4\,\AA{} is the Ni-Na atomic distance, which was not included in any of the fitting. 
			(c) Short and long Ni-O bond lengths obtained from fitting the \{2,2\} and \{2,1\} models, to the EXAFS data; Ni-O lengths from fitting the \{1,1\} model are also shown. 
			(d) Figures of merit for fits performed with the \{2,0\} and \{1,0\} models, reduced $\chi^2$ and Bayesian Information Criteria (BIC), in the vicinity of the Ni-O bond only.
		} 
		\label{EXAFS_figure}
	\end{figure*}
	
	XAS at the Ni K edge was performed, with analysis focusing on the extended X-ray absorption fine structure (EXAFS) in real-space. 
	This data, $\chi(r)$, is analogous to a partial PDF of Ni-O bond lengths, but with the significant difference that there is a phase shift resulting in peaks in $\chi(r)$ being down-shifted by around 0.5\,\AA{}. 
	It is therefore a useful supplement to the PDF data. 
	The room-temperature $\chi(r)$ data presented here resembles the EXAFS on NaNiO$_2$ reported previously~\cite{jacquet2024fundamental}.
	
	Both the X-ray absorption near edge structure (XANES) and EXAFS change on heating through $T_\mathrm{JT}$ [Figure~\ref{EXAFS_figure}]. 
	In the XANES, although there is no change in peak position (indicating that the Ni oxidation state does not change from Ni$^{3+}$) there is a pre-edge feature around 8335\,eV in $d\mu/dE$ which is more prominent for the $T<T_\mathrm{JT}$ regime than the $T>T_\mathrm{JT}$ regime.
	This pre-edge is likely due to quadrupole forbidden transitions, and will be less prominent in higher symmetry Ni environments. This observation is therefore consistent with the absence of a local Jahn--Teller distortion at $T>T_\mathrm{JT}$. 
	
	The EXAFS data, Figure~\ref{EXAFS_figure}(b), also shows a significant change on heating. 
	The two peaks corresponding to two Ni-Ni interatomic distances at $\sim$2.5\,\AA{} when $T<T_\mathrm{JT}$ merge into a single peak in the $T>T_\mathrm{JT}$ regime, consistent with the single Ni-Ni interatomic distance in the rhombohedral structure. 
	The peak at $\sim$1.5\,\AA{} corresponds to the Ni-O bond, which exhibits a far more subtle change in shape through the Jahn--Teller transition than the Ni-Ni peak, and fitting to the data is required to examine the changes in local structure. 
	
	Models were obtained using FEFF~\cite{newville2001exafs} and refined against the $\chi(r)$ data over different ranges. 
	Two different R(\AA{}) ranges were considered; one with just the Ni-O shell R(\AA{}) = 0.5\ -- 2.0\,\AA{}, and the other R(\AA{}) = 0.5\ -- 3.1\,\AA{} to also include the Ni-Ni interatomic distances. 
	Models with both 1 or 2 Ni-O and Ni-Ni distances were considered and we describe these using curly braces enclosing the comma-separated values for the number of lengths for each bond; for instance a model with a single Ni-O distance in the reduced range is described as \{1,0\}, whereas a model with two Ni-O bond lengths and two Ni-Ni bond lengths is described as \{2,2\}. 
	Five models were used, over both ranges: \{1,0\}, \{2,0\}, \{1,1\}, \{2,1\}, and \{2,2\}. 
	Models in which the first number is 1 correspond to JT-undistorted, and models where the first element is 2 are compatible with a JT distortion.
	
	The data in the Ni-O range (0.5\,\AA{} to 2.0\,\AA{}) were fitted well with both the JT-distorted \{2,0\} and JT-undistorted \{1,0\} models [Figure~S32] at all studied temperatures, with the \{2,0\} model consistently fitting better than the \{1,0\} model. 
	It is then necessary to determine whether the improved fit is due to a more realistic model or simply due to the additional number of refined parameters with the JT-distorted \{2,0\} model. 
	We consider two figures of merit for evaluating fitting quality, the reduced $\chi^2$ (r$\chi^2$) and the Bayesian Information Criterion (BIC)~\cite{neath2012bayesian}. 
	The $T$-dependence of these parameters, Figure~\ref{EXAFS_figure}(d), indicates that, in the monoclinic regime, model \{2,0\} better describes the data (consistent with the established picture of static cooperative JT-distorted NiO$_6$ octahedra), but when $T>T_\mathrm{JT}$ the model \{2,0\} only achieves a better fit as it has more refined parameters; i.e., the JT-undistorted model \{1,0\} is the better description of the Ni-O shell. In addition, at $T>T_\mathrm{JT}$ the short and long Ni-O bond lengths converge within error for the \{2,0\} model [Figure~S33].
	
	Over the larger range (0.5\,\AA{} to 3.1\,\AA{}) fitting was performed using models \{2,2\}, \{2,1\}, and \{1,1\}. 
	When $T<T_\mathrm{JT}$, the \{1,1\} and \{2,1\} models perform poorly as they cannot reproduce the Ni-Ni splitting observed experimentally, Figure~\ref{EXAFS_figure}(b). 
	The temperature-dependence of the figures of merit, Figure~S30, indicate that below $T_\mathrm{JT}$, \{2,2\} is the favoured model, consistent with diffraction. 
	Above $T_\mathrm{JT}$, all three models qualitatively fit the data well, but the two figures of merit support different models; BIC favours the undistorted model \{1,1\} whereas $r\chi^2$ the \{2,2\} model. 
	
	Does the EXAFS data support the possibility of a local Jahn--Teller distortion in NaNiO$_2$ above $T_\mathrm{JT}$? 
	This seems unlikely for the following reasons. 
	Firstly, when the fit is restricted solely to the Ni-O bonds the JT-undistorted model \{1,0\} fits best according to both metrics. 
	Secondly, the Ni-O bond length separation for the \{2,2\} model ($\sim$0.15\,\AA{}, Figure~\ref{EXAFS_figure}(c)), are significantly reduced compared to the separation at $T<T_\mathrm{JT}$ (around 0.27\,\AA{}), which differs from the order-disorder transition in LaMnO$_3$ where there is little change in bond length from EXAFS~\cite{souza2005local}. 
	Thirdly, when we plot histograms of the four shortest and two longest Ni-O bonds in the big box PDF data at 500\,K, Figure~S10, which is analogous to the \{2,...\} models used for the EXAFS fitting, we also observe a bond length separation despite other forms of analysis showing the absence of Jahn--Teller distortions according to the PDF data; this suggests that the \{2,2\} EXAFS model is consistent with a thermally-disordered, JT-undistorted state above $T_\mathrm{JT}$. 
	For these reasons, even if the \{2,2\} model is favoured by EXAFS above $T_\mathrm{JT}$, this could be explained without requiring a local Jahn--Teller distortion. 
	We therefore interpret the EXAFS data as consistent with the absence of local Jahn--Teller distortions for $T>T_\mathrm{JT}$.

	\subsection{\textit{Ab initio} molecular dynamics.}
	
	\begin{figure*}[t]
		\includegraphics[scale=0.42]{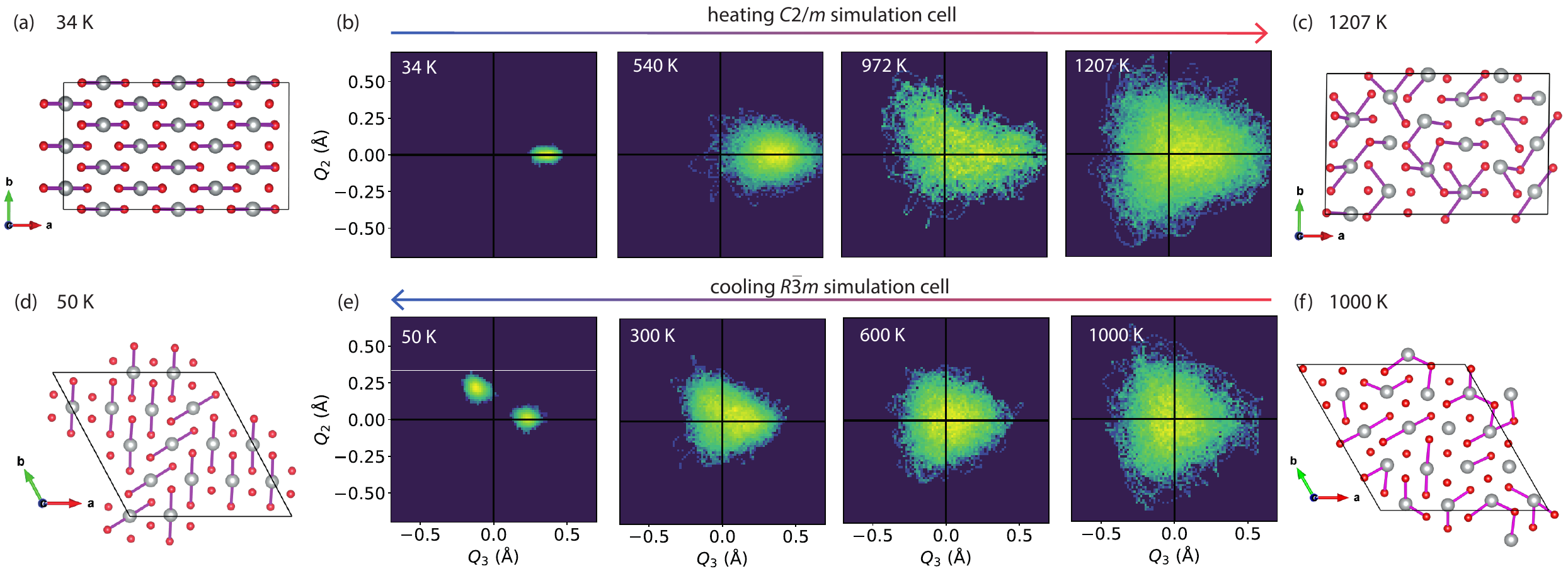}
		\caption{
			(a,c,d,f) Snapshots of a NiO$_6$ layer as obtained from AIMD simulations with $C2/m$ (a,c) and $R\bar{3}m$ (d,f) starting configurations, at various temperatures. Ni: grey; O: red. Ni-O bonds above 2.1\,\AA{} are shown. 
			(b,e) Van Vleck $E_g(Q_2,Q_3)$ diagrams as a function of temperature for both configurations, showing the displacive transition; note that a rotation of 120$^\circ$ from the $Q_2=0$ line means a change in the axis of elongation. 
			Analogous figures based on the PDF big box analysis can be seen in Figure~S11. 
		} 
		\label{AIMD-snapshots}
	\end{figure*}
	
	\begin{figure*}[t]
		\includegraphics[scale=0.85]{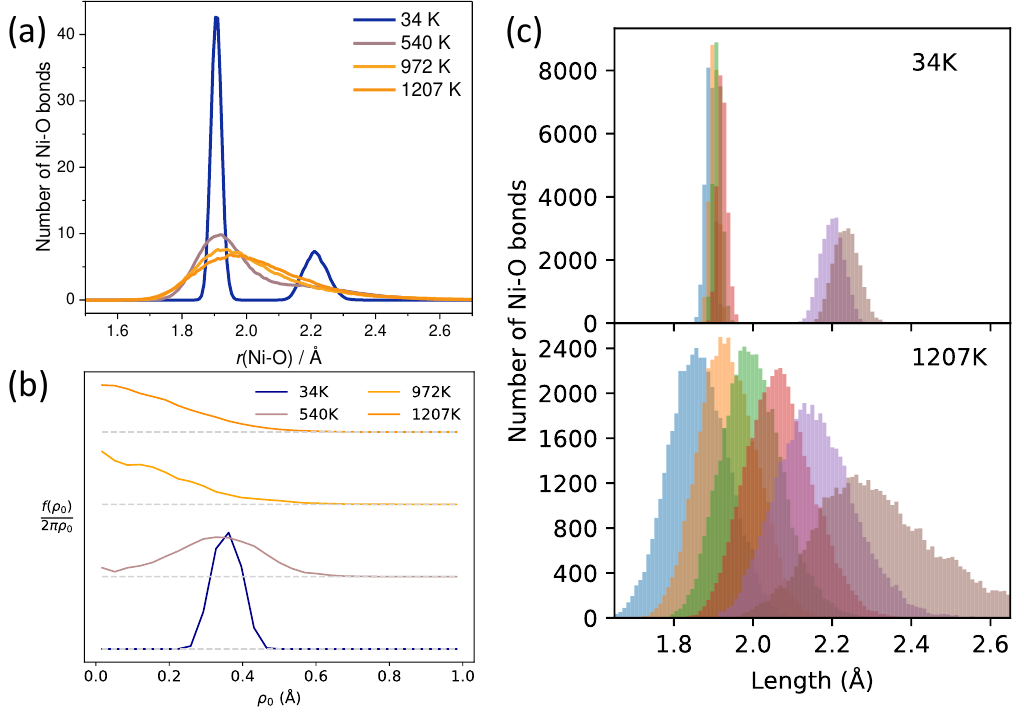}
		\caption{
			(a) Calculated Ni-O pair distribution functions from AIMD; these are convolved with experimental $Q_\mathrm{max}$ in Figure~S36. 
			(b) The probability function of $\rho_0$  [Equation~\ref{rho_0_prob}] from AIMD, for the cells with the $C2/m$ starting structure. 
			(c) Ni-O bond lengths at two temperatures for AIMD runs from the $C2/m$ starting structure.
		}
		\label{AIMD_figure}
	\end{figure*}
	
	\textit{Ab initio} molecular dynamics simulations of NaNiO$_2$ simulation cells starting with a collinear ordering of Jahn--Teller distortions, in a $3\times3\times3$ supercell (216 ions) expanded from the monoclinic cell, were performed at temperatures between 34 and 1207\,K [Figure~\ref{AIMD-snapshots}(a,c)]. 
	Both the cell shape and volume were allowed to change. 
	At low temperatures, the AIMD trajectories show thermal vibrations but the JT distortions persist, and remain collinear. 
	On heating, the distortions decrease in magnitude and are no longer aligned to a single axis. 
	At high temperatures, the octahedra approach a JT-undistorted state. 
	
	The temperature-dependence of the Jahn--Teller distortions is quantified by looking at the Ni-O PDF [Figure~\ref{AIMD_figure}(a)] and the probability function of $\rho_0$ [Figure~\ref{AIMD_figure}(b)]. 
	At low temperatures (34\,K), the PDF shows distinct peaks corresponding to short and long Ni-O bonds with a large, positive $\rho_0$. 
	On heating (972\,K, 1207\,K) the two peaks in the PDF merge to a single broad asymmetric peak. 
	$\rho_0$ significantly decreases, with the highest probability at $\rho_0=0$. 
	This can also be seen in the heat map distributions in $E_g(Q_2,Q_3)$ shown in Figure~\ref{AIMD-snapshots}.

	However, at these high temperatures there remains a residual distortion evident from a small asymmetry in the Ni-O PDF distribution, with a preference for elongation along the collinear axis [Figure \ref{AIMD-snapshots}(b) 1207\,K]. 
	We attribute this to the finite simulation cell size enforcing a periodicity in the structure that may cause longer-range correlations of the NiO$_6$ octahedra. 
	To explore this, simulations were run from a similarly-sized starting configuration ($4\times4\times1$ expansion of the rhombohedral unit cell; 192 ions) without JT distortions, and hence without inter-octahedral correlations to begin with, cooling the sample from 1000\,K. 
	The same underlying behaviour is observed as seen in the simulations starting from JT distorted simulation cells, i.e. distortions are observed at $T<T_\mathrm{JT}$ that vanish at $T>T_\mathrm{JT}$ and (re)emerge at $T<T_\mathrm{JT}$. 
	At \textit{ca.} 1000\,K, the NiO$_6$ octahedra are fully isotropic [Figure~\ref{AIMD-snapshots}(e) 1000\,K]. On cooling, Jahn--Teller distortions emerge. 
	These are row-ordered [Figure~\ref{AIMD-snapshots}(e) 300\,K], i.e. arranged collinearly within a row, rather than across the simulation cell, as predicted previously~\cite{radin2018simulating}. 
	The rows show a preference for two out of the three possible directions of the distortions (i.e. $\arctan{(Q_2/Q_3)} = 0$ or $2\pi/3$ but never $4\pi/3$), resulting in two clusters of data points in the van Vleck plot at low temperatures [Figure~\ref{AIMD-snapshots}(e) 50\,K]. 
	This is likely a consequence of the very small domain sizes, resulting presumably from the rapid cooling on the timescale of the AIMD simulations (a few picoseconds), not allowing the system to arrange into a macroscopically collinear ground state. 
	We note that in the transition regime for both simulation cells, the octahedral distortions are elongated along two out of the three octahedral axes [Figure~\ref{AIMD-snapshots}(b) 972\,K,(e) 300\,K], suggesting the transitions between the collinearly distorted phase with elongations along one axis and the displacive phase with random fluctuations along all axes may occur through an intermediate state with distortions occurring preferentially along two axes. If cooled slowly, the distortions are expected to order collinearly, whereas the rapid quenching of the AIMD simulation freezes the distortions in. 
	
	The calculated transition temperature between the JT-distorted state and the JT-undistorted state depends on the initial configuration of the simulation cell, and this does not change when the length of the AIMD runs is extended further. 
	On heating the distorted simulation cell, the onset of the JT-transition $T_\mathrm{onset}$ is found to occur at around 586\,K, i.e. at a slightly higher temperature than the experimental onset temperature obtained from synchrotron X-ray diffraction, 460\,K.
	On cooling the undistorted simulation cell, the onset of the transition $T_\mathrm{onset}$ is below 300\,K, i.e. slightly lower than the experimental onset temperature.
	A closer look at these differences in $T_\mathrm{onset}$ reveals that there are two factors causing the deviations. 
	First, as the transition is a first-order transition, hysteric behaviour is expected, i.e. the transition temperature is expected to differ between heating and cooling, and, albeit to a smaller extent, hysteresis has been observed experimentally, too.~\cite{chappel2000study,sofin2005new}
	Second, the AIMD simulations starting from the distorted and undistorted simulation cells result in different domain sizes, affecting the transition temperatures further. 
	The cooperative collinear distortions of the distorted simulation cell correspond to a scenario of infinitely large domains which shift the transition temperature to its maximum value. When cooling the undistorted simulation cell, very small domains form [Figure~\ref{AIMD-snapshots}(e) 300\,K] resulting in short-range correlations between octahedral distortions.
	The transition temperature obtained from cooling the undistorted simulation cell therefore constitutes a lower cutoff to the transition temperature from the JT-undistorted phase to the distorted phase. 
	The AIMD simulations thus predict the transition from the JT-distorted to the JT-undistorted phase to occur at $T<586$\,K, and the transition from the JT-undistorted to the JT-distorted phase at temperatures $T>300$\,K. 
	The onset temperature derived for the transition to the JT-undistorted phase based on synchrotron XRD of samples of intermediate domain sizes in this study, 460\,K, lies approximately in the middle of the predicted temperature window.

	\section{\label{sec:level5}Discussion}
	
    Our studies of the local Ni$^{3+}$ environment, using neutron PDF and Ni K edge EXAFS, suggest the absence of local static Jahn--Teller-distortions for $T>T_\mathrm{JT}$. 
    Further, while a dynamic Jahn--Teller distortion~\cite{goodenough1998jahn} has previously been suggested for LiNiO$_2$~\cite{stoyanova1993magnetic,sugiyama2010low,sicolo2020and}, by analogy with similar proposals for LaMnO$_3$~\cite{zhou1999paramagnetic}, such a dynamic Jahn--Teller distortion would also be resolvable via our neutron PDF and EXAFS measurements. 
    These observations suggest that the JT transition in NaNiO$_2$ is better characterised as displacive rather than order-disorder~\cite{radin2020order}. 
    This is different to the observations of an order-disorder transition in LaMnO$_3$~\cite{araya2001local,souza2005local,qiu2005orbital,garcia2005jahn}, and the proposed order-disorder transition for NaNiO$_2$~\cite{radin2020order}. 
    To our knowledge, only in LiNiO$_2$~\cite{genreith2023jahn} has prior evidence for a displacive Jahn--Teller transition been put forward, using AIMD and inferences from changes in strain. 
    
    As in previous reports we observe a first order phase transition in NaNiO$_2$ with the monoclinic and rhombohedral phases coexisting over a finite temperature range. 
    The observed discontinuous increase in unit cell volume from the monoclinic to the rhombohedral structures [Figure~\ref{I11-synchrotron}] is also consistent with a first-order phase transition. 
    The positive thermal expansion through the transition is consistent, via the Clausius-Clapeyron equation, with our previous finding~\cite{nagle2022pressure} of $dT_\mathrm{JT}/dP > 0$. We note that in the displacive transition in SrTiO$_3$~\cite{fossheim1972ultrasonic} a positive $dT_\mathrm{a}/dP$ is also observed. 
    In contrast, a negative thermal expansion is found at the transition temperature in LaMnO$_3$~\cite{chatterji2003volume} and ferroelectric PbTiO$_3$~\cite{rossetti1998phase,pan2022tolerance}, both order-disorder transitions, with $dT_c/dP < 0$~\cite{zhou2008breakdown,sani2002pressure}. 
    The displacive transition may therefore be related to the sign of $dT_\mathrm{JT}/dP$. 
	
	In NaNiO$_2$, the high-temperature phase will be enthalpically unfavourable and hence the phase transition will be entropically-driven, according to the free energy $\Delta G = \Delta H - T \Delta S$, where $\Delta H$ is the enthalpy change, $T$ is temperature, and $\Delta S$ is the entropy change. 
	There are two relevant contributions to the entropy: configurational and vibrational. 
	We find the configurational entropy of orbital disorder for LiNiO$_2$ and NaNiO$_2$ to be subextensive (see Appendix A), meaning it does not increase with increasing system size~\cite{camp2012subextensive}. A subextensive configurational entropy has also been found for an Ising model on an elastic triangular lattice~\cite{shokef2011order}. 
	Consequently the entropy term is dominated by dynamic vibrational effects which disfavour local Jahn--Teller distortions. 
	This is likely a contributing factor in the displacive JT transition we have observed, and is in contrast to high-temperature LaMnO$_3$ where, by analytic and Monte-Carlo calculations using the Potts model, there is extensive configurational entropy of orbital disorder~\cite{ahmed2005phase,ahmed2006potts} and hence an order-disorder transition. 
	 
	An alternative hypothesis we considered for the disappearance of Jahn--Teller distortions in high-temperature NaNiO$_2$ was the possibility for delocalisation of the electrons resulting in a change in the energy landscape of the system. This is by analogy to the insulator-to-metal transition (IMT) previously observed in the perovskite nickelates~\cite{alonso1999charge,bisogni2016ground}. 
	However, while variable-temperature conductivity, $\sigma(T)$, data from Ref.~\cite{delmas1994effect} [Figure~S6], shows an increase in conductivity by several orders of magnitude on heating, there is no change of sign of $d\sigma(T)/dT$. 
	This is similar to the conductivity data for LaMnO$_3$~\cite{zhou1999paramagnetic}, which would suggest the electronic conductivity is not closely related to the order-disorder/displacive nature of the transition. 
	
	There are a number of other differences between NaNiO$_2$ and LaMnO$_3$ which could affect the Jahn--Teller transition. 
	For instance, the edge-sharing octahedral connectivity in NaNiO$_2$ results in much larger intra-octahedral angular distortion compared with the corner-sharing octahedra in LaMnO$_3$. 
	This difference in connectivity is a consequence of the triangular Ni network in NaNiO$_2$. 
	Given the constraint that axes of elongation of adjacent JT-distorted Ni atoms cannot point at the same O anion, there are no possible long-range JT-disordered configurations~\cite{chung2005local}, which means that this triangular Ni network cannot host a Potts-type Jahn--Teller disorder such as that proposed for LaMnO$_3$~\cite{ahmed2005phase,ahmed2006potts,thygesen2017local}. 
	Another difference is that $d^4$ Mn$^{3+}$ is generally found to be JT-distorted, whereas nominally $d^7$ Ni$^{3+}$ is susceptible to other types of electronic configuration such as charge disproportionation~\cite{garcia1994neutron,mizokawa2000spin,wawrzynska2007orbital,kang2007valence,garcia2009structure}; this may be due to the smaller electronegativity difference between Ni-O, compared with Mn-O, which has been argued to cause charge-transfer insulating~\cite{zaanen1985band} behaviour in many nickelates~\cite{ronda1987photoconductivity,bisogni2016ground,foyevtsova2019band}. 
	Attempting to deconvolute the role of these different factors will require further study on the transition in other Jahn--Teller-distorted materials, along with theoretical and computational work. 
	
	Whether or not a JT transition is displacive or order-disorder may be important in some battery cathode materials. This will be particularly apparent if the JT transition occurs below room-temperature, as it does in, for instance, the NaNi$_x$Co$_{1-x}$O$_2$ solid solution for $x\le0.8$~\cite{delmas1994effect} and the spinel LiMn$_2$O$_4$~\cite{yamada1995jahn}. In these cases, the presence or absence of local JT distortions will result in different ionic mobility, and will impact the crystallographic changes that occur with electrochemical cycling. The implications of this on electrochemistry should be studied in future works. 

	\section{\label{sec:level6}Conclusion}
	
	We have shown that the Jahn--Teller transition in NaNiO$_2$ is displacive in character, with the absence of local JT-distortions or orbital ordering in the NiO$_6$ octahedra at $T>T_\mathrm{JT}$ ($T_\mathrm{JT}$ being the cooperative Jahn--Teller transition temperature). 
	To our knowledge this is the first time that local probes of structure have shown a displacive Jahn--Teller transition, though order-disorder transitions have been reported for other JT-distorted perovskite LaMnO$_3$~\cite{araya2001local,garcia2005jahn,souza2005local,qiu2005orbital}.
	
	Our findings are not consistent with previous computational studies on NaNiO$_2$, which predicted an order-disorder transition with persisting local JT distortions~\cite{radin2020order}, but our own \textit{ab initio} molecular dynamics calculations support our experimental findings. This complementarity between local probe experiments and simulation provides further support for our earlier \textit{ab initio} molecular dynamics study on LiNiO$_2$ which also predicted a displacive JT transition~\cite{genreith2023jahn}.
	
    We found that the configurational entropy of orbital disorder is subextensive in layered transition metal oxides such as NaNiO$_2$, and contrast this to the extensive configurational entropy~\cite{ahmed2005phase} in the case of perovskites such as LaMnO$_3$. 
    Further theoretical work is required to better understand these different behaviours. 
	
	\section*{\label{sec:level2}Methods}

	\subsection{Sample synthesis.} 
	
	Samples were prepared by solid state synthesis. Na$_2$O$_2$ (Alfa Aesar; 95\%) and NiO (Alfa Aesar; 99.995\%) were mixed and pelletised in a 1.05:1 molar ratio of Na:Ni, with excess Na to account for Na-loss during heating. 
	The pelletised precursor mixture was placed in an alumina crucible and heated to 973\,K for 70\,hrs in a tube furnace under a constant flow of O$_2$. O$_2$ was maintained throughout the heating and cooling processes. 
	The sample showed a colour change from light green (the NiO/Na$_2$O$_2$ precursor mixture) to black, indicating successful synthesis. 
	To prevent reaction with moisture, the sample was stored and handled in an inert Ar-atmosphere. 
	The same sample was previously studied in Ref.~\cite{nagle2022pressure}, where electron microscopy and Rietveld~\cite{rietveld1969profile} analysis of laboratory X-ray diffraction and neutron diffraction data are presented.
	
	\subsection{Synchrotron X-ray diffraction.}
	
	Variable-temperature, ambient-pressure synchrotron X-ray diffraction was performed using the I11 instrument~\cite{thompson2009beamline,thompson2011fast} at Diamond Light Source ($\lambda=0.824110$\,\AA{}) using the Mythen-2 position-sensitive detector, with a data collection time of $\sim$10\,seconds. 
	The sample was contained in a 0.5\,mm diameter glass capillary, sealed inside a glovebox with epoxy (Loctite Double Bubble). 
	The sample was heated at a rate of 12\,K/min from 322\,K to 796\,K, with periodic measurements. 
	
	Data were analysed by sequential Rietveld refinement~\cite{rietveld1969profile} using the software package \textsc{Topas 7}~\cite{coelho2018topas}. 
	A pseudo-Voigt peak function was used, and background was fitted using a Chebyshev polynomial with 20 terms. 
	A small correction for preferred orientation was applied, using the March-Dollase model~\cite{march1932mathematische,dollase1986correction}. 
	
	\subsection{Nuclear magnetic resonance.}
	
	Samples for variable-temperature nuclear magnetic resonance (VT-NMR) were packed into 4\,mm ZrO$_2$ magic angle spinning (MAS) rotors in an Ar filled glovebox, and fitted with ZrO$_2$ caps. 
	$^{23}$Na chemical shifts were calibrated using solid NaCl as an external secondary reference (7.21\,ppm relative to 1 M NaCl(aq) at 0.0\,ppm). 
	$^{23}$Na MAS NMR spectra were acquired using a Bruker Avance IIIHD spectrometer (v$_0$[$^1$H] = 500.13\,MHz, v$_0$[$^{23}$Na] = 132.46\,MHz, v$_0$ [$^{207}$Pb] = 104.26\,MHz) spectrometer, with a Bruker 4\,mm HX probe and 14\,kHz MAS. 
	Projection magic-angle turning and a phase-adjusted sideband separation (pj-MATPASS) pulse sequence was used to obtain the isotropic $^{23}$Na spectra~\cite{hung2012isotropic}.
	
	Static VT-NMR samples were measured in the same spectrometer using a Bruker HX static probe. 
	A Hahn-echo pulse sequence with $\pi/4$ = 2.05\,$\mu$s optimized on solid NaNiO$_2$ was used. 
	Temperature was calibrated through a NMR shift thermometer compound Pb(NO$_3$)$_2$, based on the known temperature-dependence of the isotropic chemical shift of Pb(NO$_3$)$_2$~\cite{takahashi1999207pb}. 
	$^{207}$Pb NMR spectra were acquired with the same probe at the same temperature values, using a Hahn-echo pulse sequence. 
	The isotropic chemical shift values for Pb(NO$_3$)$_2$ were obtained by fitting the data using the \textsc{SOLA} fitting program within \textsc{Topspin 4.1.4}.
	
	\subsection{Neutron total scattering.}
	
	Variable-temperature total-scattering neutron diffraction was performed on the NOMAD instrument~\cite{neuefeind2012nanoscale} at the Spallation Neutron Source, Oak Ridge National Laboratory, USA. 
	NaNiO$_2$ was stored under Ar in a sealed NMR tube (2\,cm sample height, 5\,mm outer diameter) for the measurements. 
	Heating was performed using a furnace, at a rate of 5$^\circ$K per minute. The sample was measured on heating at 293\,K, 450\,K, 500\,K, and after cooling at 316\,K. The neutron diffraction data and the refined crystal structure was previously published in Ref.~\cite{nagle2022pressure}. 
	In the Fourier transform to real space, a sliding $Q_\mathrm{max}$ as a function of $r$ is used, with $Q_\mathrm{max}=40$\,\AA$^{-1}$ at low-$r$ and gradually decreasing $Q_\mathrm{max}$ beyond that, with $Q_\mathrm{max}=20$\,\AA$^{-1}$ above $r=25$\,\AA{}.
	
	\subsection{Analysis of pair distribution function.}
	
	Small-box pair distribution function analysis, also known as real-space Rietveld refinement, was performed using \textsc{Topas 6}~\cite{coelho2018topas}. 
	Data were fitted in the range 1.0\,\AA{} to 10\,\AA{} in real space only. 
	The starting structures were the normal $C2/m$ and $R\bar{3}m$ structures used commonly in the literature for the low- and high-temperature phases of NaNiO$_2$.
	
	Pair distribution function analysis was performed via a ``big box" approach used \textsc{Topas 7}~\cite{coelho2018topas}, following a modified version of the method introduced in Ref.~\cite{fuller2020oxide}. 
	A supercell model was refined against both the Bragg data from NOMAD bank 4 (as defined in Ref.~\cite{neuefeind2012nanoscale}) and the experimental neutron pair distribution function. 
	In this refinement, atomic coordinates were refined until convergence following the Rietveld algorithm~\cite{rietveld1969profile}. 
	During initial iterations, restraints were applied to prevent large atomic movements, but these were removed in later stages of the refinement. 
	Upon convergence, each atom was randomly shifted by a distance less than or equal to 0.1\% of the unit cell size ($\sim$0.045\,\AA{}) in each spatial dimension to help minimisation. 
	This typically occurred several hundred times. 
	The final configuration was the converged structure with the best fit quality. Our approach differs from the more commonly-used approach in \textsc{RMCProfile}~\cite{tucker2007rmcprofile}. 
	Data were fitted in the range 0.5\,\AA{} to 20\,\AA{} in real space. 
	
	A $16\times9\times3$ supercell of a pseudo-orthorhombic cell (space group: $P1$; $a\approx2.846 $\,\AA{}; $b \approx 5.321$\,\AA{}; $c \approx 15.703$\,\AA{}; $\beta=\gamma=90^\circ$; $\alpha=90^\circ$ for the 500\,K supercell and $\sim$88$^\circ$ for the $T<T_\mathrm{JT}$ supercells) was used for the refinements. 
	This NaNiO$_2$ unit cell was obtained via transformation from the rhombohedral cell using Equation~S5 in Supplementary Information. 
	The supercell has edge lengths $\sim$45\,\AA{} and contains 10368 atoms. 
	In this analysis, the thermal parameters for all atoms were fixed at low values ($B_\mathrm{iso}=0.01$\,\AA{}$^2$) such that thermal effects are modelled by the deviation of each site from its ideal position. 
	To ensure that chemically reasonable models were obtained, soft restraints were included. The relative weighting of restraints against data was set to ensure refinements were not dominated by restraints. 
	The SI includes refinements performed in the absence of restraints to check their influence on the conclusions (Section~S3.9).
	The main soft restraint was the calculated bond valence sum~\cite{brown1985bond} (as an analog for oxidation state), $V_\mathrm{calc}$, of cations deviates from its expected value (Na: +1; Ni: +3), an established approach~\cite{norberg2009bond} in PDF analysis~\cite{marrocchelli2009cation,abrahams2010combined,liu2011neutron,norberg2011comparison,norberg2011structural,norberg2011neutron,burbano2012oxygen,chong2012local,norberg2012pyrochlore,leszczynska2013thermal,norberg2013proton,keeble2013bifurcated,payne2013fluorite,kalland2016c,chen2017role,diaz2018local,borowska2018local,kitamura2019local,kitamura2019study,fuller2020oxide,marlton2021broad,krayzman2022incommensurate,miyazaki2022reverse,ming2024dopant}. 
	Here, $V_\mathrm{calc}$ is calculated as follows:
	
	\begin{equation}\label{V_eq}
	V_\mathrm{calc} = \sum_{i=1}^6 \exp{\left[\frac{r_0-r_i}{B}\right]}
	\end{equation}
	where the empirical parameters $r_0$ and $B$ are based on the cation and anion species, and $r_i$ are the bond lengths between the metal atom and the $i$th oxygen. 
	In this work, $B=0.37$\,\AA{} and $r_0^\mathrm{Na} = 1.672$\,\AA{} and $r_0^\mathrm{Ni} = 1.7335$\,\AA{}; these values were selected to give average bond lengths of 2.335\,\AA{} and 1.99\,\AA{} for NaO$_6$ and NiO$_6$ octahedra respectively, to match the room-temperature average structure. 
	
	Since the $P1$ space group has a floating origin, a restraint was applied to keep the average shift of Ni sites close to zero.
	
	The supercells obtained from refinement were then analysed using \textsc{VanVleckCalculator}~\cite{naglecocco2023van,VanVleckCalculator}, a Python 3~\cite{10.5555/1593511} code which is based on \textsc{PyMatGen}~\cite{ong2013python}.
	
	\subsection{Ni K edge X-ray absorption spectroscopy.}
	
	X-ray absorption spectroscopy (XAS) measurements were performed on the BM23 instrument~\cite{mathon2015time} of the European Synchrotron Radiation Facility (ESRF), France. 
	Powdered NaNiO$_2$ was mixed homogeneously with a powdered boron nitride binder ($\sim$16\,mg:$\sim$70\,mg NaNiO$_2$:BN) and pressed into a 13\,mm pellet in a dry, oxygen-free nitrogen environment. 
	The pellet was cut and placed in a sealed sample holder under flowing helium. Temperature control was achieved using a resistive heater, and at each temperature the sample was left for 6 minutes to thermally equilibriate. 
	At each temperature the sample was measured ten times to ensure reproducibility, and an average spectrum produced. X-ray absorption was measured in the vicinity of the Ni K edge ($\sim8.3$\,keV).
	
	X-ray absorption was calculated using the Beer-Lambert law~\cite{kerr2022characterization} comparing X-ray intensity in ionisation chambers before and after transmission through the sample, and a third ionisation chamber measured the absorption through Ni foil as a reference. 
	Data normalisation was performed using the Python-based~\cite{10.5555/1593511} \textsc{Larch}~\cite{newville2013larch} package. 
	For the normalisation of the data, a pre-edge range between -350\,eV and -45\,eV from the edge centre was fit with a straight line, and the post-edge region between 200\,eV and 1250\,eV was fit with a quadratic polynomial. 
	Data were transformed from $E$-space to $k$-space via the transform $k = \sqrt{\frac{2m_e}{\hbar^2} \left(\hbar \omega - E_0 \right)}$, where $m_e$ is the mass of the electron, $E_0$ is the edge energy, and $\omega$ is the frequency of the measured photon. 
	Finally, the EXAFS function $\chi(k)$ was obtained by fitting and subtracting a spline to the EXAFS data in $k$-space, where $R_\mathrm{bkg}$ was set to 1\,\AA{} and the spline was fit within a $k$-range of 0\,\AA{}$^{-1}$ to 18.25\,\AA{}$^{-1}$, with a $k^2$-weighting. 
	For the Fourier transform of EXAFS from $k$-space to $R$-space, a $k$-range of 2.5\,\AA{}$^{-1}$ to 15\,\AA{}$^{-1}$ was used, and to minimise Fourier ripples from the cutoffs in $k$-space, a Kaiser-Bessel window with end width $dk=7$\,\AA{}$^{-1}$ was applied to the data in $k$-space before the transformation.
	
	Calculated EXAFS data were obtained by multiple-scattering path expansion using FEFF~\cite{newville2001exafs}, as implemented in Larch in Python. In this model, data is fit using the EXAFS equation over each set of equivalent paths $j$, defined as follows:
	
	\begin{equation}
		\chi (k) = S_0^2 \sum_j \frac{N_j |f_{j}(k)|}{kR_j^2} \sin{\left[2kR_j + \phi_j(k)\right]} \\
		\times \exp{\left[-\frac{2R_j}{\lambda_j(k)}\right]} \exp(-\sigma_j^2 k^2)
	\end{equation}
	where FEFF is used to calculate scattering amplitude $|f_{j}(k)|$, photoelectron mean free path $\lambda_j(k)$, and phase $\phi_j(k)$ for all paths. $N_j$ is the number of equivalent paths for each $j$, which is fixed. 
	The amplitude reduction factor $S_0^2$ was kept fixed at a value obtained by fitting against the Ni foil reference, $S_0^2=0.79286024$. 
	The path distance $R_j$ and Debye-Waller factors $\sigma_j$ for each path were refined during model fitting. 
	Edge position $E_0$ was also freely refined.
	
	\subsection{\textit{Ab initio} molecular dynamics.} 
	
	AIMD simulations were performed according to the Generalized Gradient Approximation (GGA)~\cite{perdew1996generalized} and the projector augmented wave method (PAW)~\cite{blochl1994projector}, as implemented in the Vienna Ab Initio Simulation Package (VASP)~\cite{kresse1996efficient,kresse1999ultrasoft}. 
	The plane-wave energy cutoff was set to 500\,eV. 
	Supercells are expansions of the monoclinic ($3\times3\times3$; 216 ions) and rhombohedral ($4\times4\times1$; 192 ions) unit cells, respectively. 
	Simulations were performed at the $\Gamma$ point and checked against calculations with a $2 \times 2 \times 2$ Monkhorst-Pack~\cite{monkhorst1976special} \textit{k}-point grid for consistency. 
	The convergence criteria for the electronic and ionic relaxations were set to $10^{-6}$\,eV and $5\cdot10^{-3}$\,eV/\AA{}, respectively.
	
	For Ni, the $4s^23d^8$ electrons were treated as valence electrons. 
	To account for the strongly correlated d electrons, a rotationally invariant Hubbard \textit{U} parameter~\cite{dudarev1998electron} of $U_\mathrm{eff} = 6$\,eV was selected, which was used successfully in previous studies of layered oxide cathodes, both for 0\,K DFT calculations and finite temperature AIMD simulations~\cite{das2017first,genreith2023oxygen,genreith2023jahn}. 
	For oxygen, the $2s^22p^4$ electrons were considered as valence electrons.
	
	AIMD simulations were performed for the isothermal-isobaric ensemble (\textit{NpT}, constant pressure, particle number, and temperature) at zero pressure, allowing the cell shape and volume to change. 
	A Langevin thermostat was used with friction coefficients set to zero to minimise impact on the lattice vibrations. 
	Trajectories were run for 3\,ps at timesteps of 1\,fs. 
	Following an equilibration period of 1\,ps, snapshots were sampled every 2-5\,fs, using Ovito~\cite{stukowski2009visualization}, for the analysis of the distortions.
	
	\section*{Appendix A: proof of subextensive configurational entropy for orbital disorder}
	\renewcommand{\theequation}{A.\arabic{equation}}
	\setcounter{equation}{0}
	
	Here, a mathematical proof is presented that, given a set of assumptions, there is subextensive configurational entropy of orbital disorder in layered oxide materials based on a triangular array of JT-active cations. 
	We define subextensive as meaning that the configurational entropy does not increase with increasing system size. 
	
	This relies on three basic axioms:
	\begin{enumerate}
		\item The effect on orbital ordering of inter-layer interactions is negligible.
		\item Each oxygen anion participates in precisely one Jahn--Teller-elongated bond.
		\item Jahn--Teller distortions are purely of the $Q_3$ tetragonal elongation type.
	\end{enumerate}
	
	Axiom (1) is justified because the O-$A$-O ($A$=Li,Na) layers prevent orbital interactions because the Na does not overly constrain the position of the O atom and there is not much correlation between the O atoms around an Na site. 
	Axiom (2) because participation in more than one JT-elongated bond would make the O anion under-bonded. 
	Axiom (3) because this is how octahedra are at room-temperature in Na$M$O$_2$ ($M$=Ni,Mn) materials.
	
	Configurational entropy per layer is given by:
	
	\begin{equation}
		S_\mathrm{config}^\mathrm{layer} = \ln{\left[W_\mathrm{layer}\right]}
	\end{equation}
	where $W_\mathrm{layer}$ is the number of configurations within the layer. Note we define entropy in units of Boltzmann's constant $k_\mathrm{B}$.
	
	Given Axiom 2, there are only two possible alignments of the elongated axes between neighbouring octahedra. 
	Either the axes of elongation are parallel (``A") or angled 120$^\circ$ apart (``B"). There are three rows of Ni sites within a layer, separated by an angle 120$^\circ$. 
	In a monoclinic cell such as in NaNiO$_2$ at $T<T_\mathrm{JT}$, all interactions would be of the ``A" type. 
	If there were true orbital disorder, each chain would be a random combination of these two possible configurations ``A" and ``B", for instance "AABBABAAABAABBABABB". 
	If \textit{L} is the length of such a chain (divided by Ni-Ni distance), then we have the following number of configurations:
	
	\begin{equation}
		W_\mathrm{layer} = 3 \times 2^L = 3 \times 2^{A^{1/2}}
	\end{equation}
	where the $3$ factor occurs because there are three directions where chains occur, and we relate $L$ to an arbitrary area $A$ via $L=A^{1/2}$. 
	
	Given that $N$ is proportional to $A$, where $N$ is the number of Jahn--Teller-elongated axes within an area of the single layer, we can use $N=\alpha A$ and substitute this into the equation for configurational entropy as follows:
	
	\begin{equation}
		S_\mathrm{config}^\mathrm{layer} = \ln{ \left[3 \times 2^{\alpha'N^{1/2}} \right] } =  C + N^{1/2} \ln{\left[2\right]}
	\end{equation}
	where we have used $\alpha'=\alpha^{1/2}$ and used constant {$C = \ln{\left[3\right]} + \alpha' \ln{\left[2\right]}$}.
	
	If we then calculate configurational entropy per NiO$_6$ octahedron for a large system, as $N \rightarrow \infty $, $C/N \rightarrow 0$ and so:
	
	\begin{equation}
		\frac{S_\mathrm{config}^\mathrm{layer}}{N} \approx N^{-1/2} \ln{\left[2\right]}
	\end{equation}
	
	Therefore, as $N \rightarrow \infty $, so does $\frac{S_\mathrm{config}^\mathrm{layer}}{N} \rightarrow 0$. This indicates that the configurational entropy of orbital disorder is subextensive.
	
	\section*{Data availability}
	
	Data from the BM23 instrument at the European Synchrotron Radiation Facility is available at doi:10.15151/ESRF-ES-962076745~\cite{doi:10.15151/ESRF-ES-962076745}. 
	All other data is available in the University of Cambridge repository at doi.org/10.17863/CAM.112349~\cite{doi.org/10.17863/CAM.112349}. 
	
	\section*{Supplementary Information}
	
	A document containing Supplementary Information to this article is available. 	
	The Supplementary Information contains: A description of the Van Vleck modes. Further analysis of supercell outputs of big box PDF analysis, and results from various different starting configurations. Expanded AIMD results. NMR data. Tabulated results of all analysis techniques. 
	
	\section*{Acknowledgements}

	The authors acknowledge Oak Ridge National Laboratory, a United States Department of Energy Office of Science User Facility, for use of the NOMAD instrument at the Spallation Neutron Source (experiment IPTS25164). 
	We acknowledge the European Synchrotron Radiation Facility for provision of beam time on BM23 (experiment CH6437). 
	We acknowledge I11 beamline at the Diamond Light Source, UK, for the synchrotron XRD measurement done under BAG proposal (the data presented in this work under CY34243; essential preliminary data from CY28349). 
	Calculations were performed using the Sulis Tier 2 HPC platform hosted by the Scientific Computing Research Technology Platform at the University of Warwick (EP/T022108/1). 

	We would like to thank Dr Euan N. Bassey, \mbox{Lucy Haddad}, Dr Anastasia Yu. Molokova, Dr Chloe C. Coates, Dr Farheen N. Sayed, Dr Gheorghe-Lucian Pășcuț, and Dr João Elias F. S. Rodrigues for useful discussions.

	Crystal structure figures were prepared using \textsc{Vesta-3}~\cite{momma2011vesta} and graphs were prepared using \textsc{MatPlotLib}~\cite{hunter2007matplotlib}.

	\section*{Funding}
	
	This work was supported by the Faraday Institution (\mbox{FIRG001}, \mbox{FIRG017}, \mbox{FIRG024}, \mbox{FIRG060}). 
	L.A.V.N-C acknowledges a scholarship \mbox{EP/R513180/1} to pursue doctoral research from the UK Engineering and Physical Sciences Research Council (EPSRC) and additional funding from the Cambridge Philosophical Society. 
	J.M.A.S. acknowledges support from the EPSRC Cambridge NanoCDT, \mbox{EP/L015978/1}. 
	A.L.G. acknowledges European Research Council (ERC) funding under grant \mbox{788144}. 
	
	\bibliography{references}

\end{document}